\documentclass{midl} 

\usepackage{mwe} 
\usepackage{graphicx}
\usepackage{amsmath}
\usepackage{amssymb}
\usepackage[utf8]{inputenc}
\DeclareMathOperator*{\argmax}{arg\,max}

\jmlrvolume{-- Under Review}
\jmlryear{2020}
\jmlrworkshop{Full Paper -- MIDL 2020 submission}
\editors{Under Review for MIDL 2020}
\usepackage{array,booktabs,arydshln,xcolor}

\title[EG-CNN]{Edge-Gated CNNs for\\ Volumetric Semantic Segmentation of Medical Images}

\midlauthor{\Name{Ali Hatamizadeh\nametag{$^{1,2}$}} \Email{ahatamiz@cs.ucla.edu}\\
\Name{Demetri  Terzopoulos\nametag{$^{1}$}} \Email{dt@cs.ucla.edu}\\
\Name{Andriy Myronenko\midlotherjointauthor\nametag{$^{2}$}} \Email{amyronenko@nvidia.com }\\
\addr $^{1}$ Computer Science Department, University of California, Los Angeles, CA, USA \\
\addr $^{2}$ NVIDIA, Santa Clara, CA, USA
}

\begin{document}

\maketitle

\begin{abstract}
Textures and edges contribute different information to image recognition. Edges and boundaries encode shape information, while textures manifest the appearance of regions. Despite the success of Convolutional Neural Networks (CNNs) in computer vision and medical image analysis applications, predominantly only texture abstractions are learned, which often leads to imprecise boundary delineations. In medical imaging, expert manual segmentation often relies on organ boundaries; for example, to manually segment a liver, a medical practitioner usually identifies edges first and subsequently fills in the segmentation mask. Motivated by these observations, we propose a plug-and-play module, dubbed Edge-Gated CNNs (EG-CNNs), that can be used with existing encoder-decoder architectures to process both edge and texture information. The EG-CNN learns to emphasize the edges in the encoder, to predict crisp boundaries by an auxiliary edge supervision, and to fuse its output with the original CNN output. We evaluate the effectiveness of the EG-CNN with various mainstream CNNs on two publicly available datasets, BraTS 19 and KiTS 19 for brain tumor and kidney semantic segmentation. We demonstrate how the addition of EG-CNN consistently improves segmentation accuracy and generalization performance.  
\end{abstract}


\section{Introduction}

Image segmentation plays an important role in medical image analysis as accurate delineation of anomalies is crucial for computer aided diagnosis and treatment planning. With the advent of deep learning,  Convolutional Neural Networks (CNNs) have been successfully adopted in various medical semantic segmentation applications~\cite{gibson2018automatic,dolz20183d,Myronenko18,zhu2019anatomynet}. In particular, the seminal U-Net architecture~\cite{Ronneberger15} demonstrated the effectiveness of down-sampling and up-sampling paths for multi-scale feature representation learning, and many encoder-decoder CNNs have since been introduced based on the same principles. Medical images contain different types of representations: edges encode shape information, while the appearance of regions is manifested by textures. As such, a single processing pipeline may lead to the loss of shape information and may result in imprecise boundary definitions. It has been empirically demonstrated~\cite{geirhos2018} that unlike the human visual system, common CNN architectures are biased towards recognizing texture content. As a result, CNN predictions often need to be post-processed \cite{kamnitsas2017efficient, hatamizadeh2019deeplesion} to compensate for the shape details that are lost during training.        

The current paradigm of processing different abstractions within a single pipeline is sub-optimal. It can be remedied by utilizing effective processing of information in a structured manner, similar to the human visual perception system. In medical imaging, radiologists usually rely on identifying the boundaries of the organ/lesion of interest as a first step in manual delineation. For instance, segmenting brain tumors from MR images would entail following lesion edges and subsequently deducing the interior region.    

\begin{figure*}[t]
  \includegraphics[width=\textwidth,]{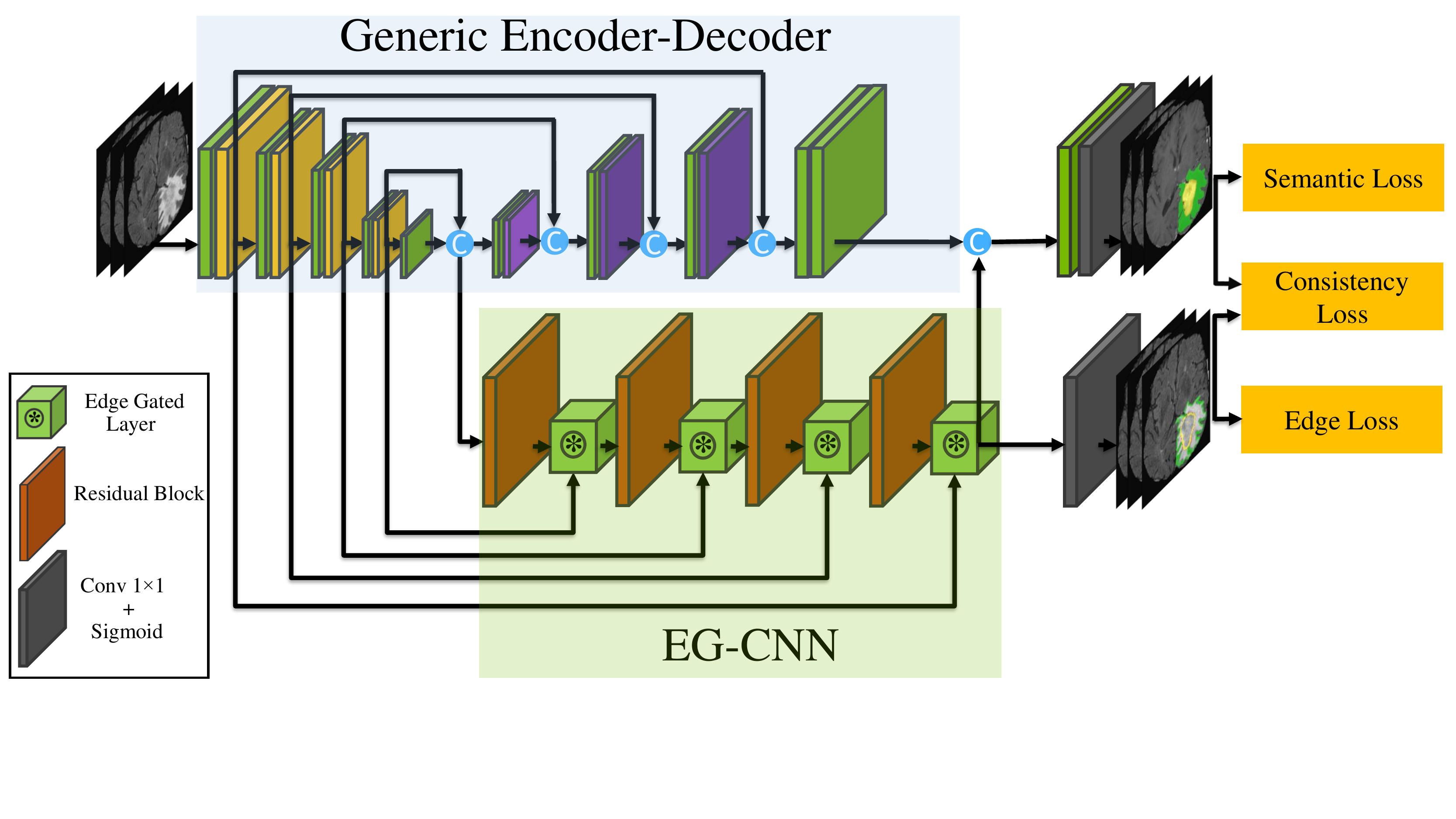}
  \caption{The proposed EG-CNN module can be integrated with any generic encoder-decoder architecture and highlight the edge representations of the intermediate feature maps.}
  \label{fig:pipeline}
\end{figure*}

Instead of proposing a new CNN architecture, we propose an novel 3D plug-and-play module that we call the Edge-Gated CNN (EG-CNN), which can be incorporated with any encoder-decoder architecture to disentangle the learning of texture and edge representations. The contribution of the proposed EG-CNN is two-fold. First, EG-CNN proposes an effective way to progressively learn to highlight the edge semantics from multiple scales of feature maps in the main encoder-decoder architecture by a novel and efficient layer denoted the edge-gated layer. Second, instead of separately supervising the edge and texture outputs, the EG-CNN introduces a dual-task learning scheme, in which these representations are jointly learned by a consistency loss. Therefore, without increasing the cost of data annotation and by exploiting the duality between edge and texture predictions, the EG-CNN improves the overall segmentation performance with highly detailed boundaries. Figure\ref{fig:pipeline} illustrates the integration of the EG-CNN with an existing encoder-decoder CNN architecture.  

We validate the effectiveness of our EG-CNN on two publicly available datasets, BraTS 2019~\cite{bakas2017segmentation} and KiTS 2019~\cite{heller2019kits19} for brain and kidney tumor segmentation, respectively. For this purpose, we utilize as backbones three popular 3D CNN architectures, U-Net \cite{Ronneberger15}, V-Net \cite{Milletari16}, and Seg-Net \cite{Myronenko18}, and our results demonstrate substantial improvement when EG-CNN is leveraged with these architectures. 

\section{Methodology}
\subsection{EG-CNN}
\label{sec:EG-CNN}

We first introduce the architecture of EG-CNN. A generic CNN encoder-decoder, as we denote the main stream, learns feature representations that span multiple resolutions. Our EG-CNN receives each of the feature maps in the main stream and learns to highlight the edge representations. In particular, the EG-CNN consists of a sequence of residual blocks followed by tailored layers, as we denote the edge-gated layers, to progressively extract the edge representations. The output of the EG-CNN is then concatenated with the output of the main stream in order to produce the final segmentation output. Furthermore, the main stream and the EG-CNN are supervised by their own dedicated loss layers as well as a consistent loss function which jointly learns the output of both streams. The edge ground-truth is generated online by applying a 3D Sobel filter to the original ground truth masks. 

Each edge-gated layer requires two inputs that originate from the main stream and the EG-CNN stream that we denote the edge stream. The intermediate feature maps from every resolution of the main stream as well as the first up-sampled feature maps in the decoder are fed to the EG-CNN as inputs. The latter is first fed into a residual block followed by bilinear upsampling before being fed into the edge-gated layer along with the input from its previous resolution in the encoder. The output of each edge-gated layer (except for the last one) is fed into another residual block followed by bilinear upsampling before being fed to the next edge-gated layer along with its corresponding input from the encoder.

\subsection{Edge-Gated Layer}

Edge-gated layers highlight the edge features and connect the feature maps learned in the main and edge streams. They receive inputs from the previous edge-gated layers as well as the main stream at its corresponding resolution.
Let $e_{r,in}$ and $m_r$ denote the inputs coming from edge and main streams, respectively, at resolution $r$. First, an  attention map, $\alpha_{r}$ is obtained by feeding each input into a $1\times1\times1$ convolutional layer, $C_{1\times1\times1}$, fusing the outputs and passing them into a rectified linear unit (ReLU)  $Re(X) = \max(0, X)$ according to
\begin{equation}
\alpha_{r}=\sigma\big(Re(C_{1\times1\times1}(e_{r,in})+C_{1\times1\times1}(m_{r}))).
\label{eq:att1}
\end{equation} 
\newline
The obtained attention map $\alpha_{r}$ is then pixel-wise multiplied by $e_{r,in}$ and fed into a residual layer with kernel $w_{r}$. Therefore, the output of each resolution in EG-CNN ,$e_{r,out}$, can be represented as
\begin{equation}
e_{r,out}= e_{r,in} \odot \alpha_{r}+e_{r,in}.
\label{eq:att2}
\end{equation} 
\newline
The computed attention map highlights the edge semantics that are embedded in the main stream feature maps. In general, there will be as many edge-gated layers as the number of different resolutions in the main encoder-decoder CNN architecture. 

\subsection{Loss Functions}

The total loss of the EG-CNN is as follows: 
\begin{equation}
L_\textrm{Tot}= L_\textrm{Semantic} + L_\textrm{Consistency}+ L_\textrm{Edge},
\label{eq:egcnnloss}
\end{equation}
where $L_\textrm{Semantic}$ represent standard loss functions used for supervising the main stream in a semantic segmentation network, $L_\textrm{Edge}$ represent tailored losses for learning the edge representations, and $L_\textrm{Consistency}$ is a dual-task loss for the joint learning of edge and texture and enforces the class consistency of predictions. 

\paragraph{Semantic Loss:}

Without loss of generality, we use the Dice loss \cite{Milletari16} for learning the semantic representations of texture according to
\begin{equation}
L_\textrm{Dice}= 1- \frac{2 \sum y_\textrm{true} y_\textrm{pred} }{\sum y_\textrm{true}^2 + \sum y_\textrm{pred}^2 + \epsilon},
\label{eq:dice}
\end{equation} 
where summation is carried over the total number of pixels, $y_\textrm{pred}$ and $y_\textrm{true}$ denote the pixel-wise semantic predictions of the main stream, and $\epsilon$ is a small constant to prevent division by zero.

\paragraph{Edge Loss:}

The edge loss used in EG-CNN comprises of Dice loss \cite{Milletari16} and balanced cross entropy \cite{yu2017casenet}, as follows:
\begin{equation}
L_\textrm{Edge}= \lambda_{1} L_\textrm{Dice} + \lambda_{2}L_\textrm{BCE},
\label{eq:attloss}
\end{equation}
where $\lambda_{1}$ and $\lambda_{2}$ are hyper-parameters. Let $e_{\textrm{pred},j}$ and $e_{\textrm{true},j}$ denote the edge prediction outputs of the EG-CNN and its corresponding groundtruth at voxel $j$, respectively. Then the balanced cross entropy $L_\textrm{BCE}$ used in (\ref{eq:attloss}) can be defined as
\begin{equation}
L_\textrm{BCE}= -\beta \sum_{j\in e_{+}} \log P(e_{\textrm{pred},j}=1|x;\theta)-(1-\beta) \sum_{j\in e_{-}} \log P(e_{\textrm{pred},j}=0|x;\theta),
\label{eq:bce}
\end{equation} 
where $x$, $\theta$, $e_{-}$, and $e_{+}$ denote the input image, CNN parameters, edge, and non-edge voxel sets, respectively, $\beta$ is the ratio of non-edge voxels to all voxels, and $P(e_{\textrm{pred},j})$ is the probability of the predicated class at voxel $j$. The cross entropy loss follows (\ref{eq:bce}) except for the fact that non-edge voxels are not weighted.

\paragraph{Consistency Loss:}

We exploit the duality of edge and texture predictions and simultaneously supervise the outputs of the edge and main stream by the consistency loss. Inspired by \cite{Takikawa2019GatedSCNNGS}, the  semantic probability predictions of the main CNN architectures and the ground truth masks are first converted into edge predictions by taking the spatial derivative in a differentiable manner as described in Section~\ref{sec:EG-CNN}. Subsequently, we penalize the mismatch between the boundary predictions of the semantic masks and the corresponding ground truth by utilizing an $L_1$ loss. Let $y_{\textrm{pred},j}$ denote the output of the main stream and $c$ represent the segmentation class. We propose a consistency loss function \begin{equation}
 L_\textrm{Consistency}= \sum_{j\in e_{+}} \bigl(\| {\nabla(\argmax(P(y_{\textrm{pred},j}=1|e;c))} \|)-\| {\nabla(y_{\textrm{true},j}} \|\bigr).
\label{eq:consistency}
\end{equation} 

Due to the non-differentiability of the $\argmax$ function, we leverage the Gumbel softmax trick \cite{jang2016categorical} to avoid blocking the error-gradient. Thus, the gradient of the $\argmax$ can be approximated according to
\begin{equation}
\frac{\partial \argmax_{t}P(y^{t})}{\partial \gamma_{j}}= \nabla_{j} \frac{e^{(\log P(y_{t})+g_{t})/\tau}}{\sum_{i} e^{(\log P(y_{i})+g_{i})/\tau}},
\label{eq:gumbeltrick}
\end{equation} 
where $\gamma$ is a differentiation dummy variable, $\tau$ is the temperature, set as a hyper-parameter, and $g_{i}$ denotes the Gumbel density function.

\section{Experiments}

\subsection{Datasets}

\paragraph{BraTS 2019:} The multimodal Brain Tumor Segmentation Challenge (BraTS 2019) serves to evaluate state-of-the-art methods for the segmentation of brain tumors by providing a 3D MRI dataset with ground truth tumor segmentation labels annotated by physicians~\cite{BratsAll2018,brats1short,brats2,brats4}. The BraTS 2019 training dataset includes 335 cases, each with four 3D MRI modalities (T1, T1c, T2 and FLAIR) rigidly aligned, resampled to $1\times1\times1$\,mm isotropic resolution and skull-stripped. The input image size is $240\times240\times155$. Annotations include 3 tumor subregions: the enhancing tumor, the peritumoral edema, and the necrotic and non-enhancing tumor core. The annotations were combined into 3 nested subregions: Whole Tumor (WT), Tumor Core (TC), and Enhancing Tumor (ET). 

\paragraph{KiTS 2019:} The Kidney Tumor Segmentation Challenge (KiTS 2019) provides data comprising multi-phase 3D CTs and voxel-wise groundtruth labels or kidneys and kidney tumors for 300 patients who underwent nephrectomy for kidney tumors between 2010 to 2018 at the University of Minnesota~\cite{heller2019kits19}. The input image size is $380\times380\times250$.

\newcommand\x{0.242}
\subsection{Implementation Details}

We implemented all the models in Pytorch.\footnote{\href{https://pytorch.org/}{https://pytorch.org/}} For our experiments with the BraTS 2019 dataset, we normalized all input images to have zero mean and unit std (based on non-zero voxels only). For the KiTS 2019 dataset, we normalized the CT data to the $[-1, 1]$ range by dividing the intensity values by 1,000 and clipping the values that fall outside this range. For training, images were resampled to $1\times1\times1$\,mm isotropic resolution and resampled back to their original resolution after the inference. For both the BraTS 2019 and KiTS 2019 datasets, all the segmentation models were trained on and tested using a 80/20 split. We used $\lambda_{1}=1$, $\lambda_{2}=0.5$ as hyperprameters in (\ref{eq:egcnnloss}) and (\ref{eq:attloss}). All models were trained on eight NVIDIA Tesla V100 16GB GPUs (DGX-1 server). We used a batch size of 8 and the Adam optimization algorithm was used with an initial learning rate of $\alpha_{0}=0.0001$ decreasing according to~\cite{myronenko2020robust}
 \begin{equation}
 \label{eq:learningrate}
 \alpha = \alpha_{0} *\left(1-e/N_{e}\right)^{0.9} 
 \end{equation}
 with epoch counter $e$ and total number of epochs $N_{e}$.

 \subsection{Evaluation Metrics} 
 
 For the BraTS 2019 challenge, we used the Dice function as a standard metric for image segmentation to asses the quality of the segmentation in the vicinity of boundaries. For the KiTS 2019 challenge, we adopted the three evaluation metrics of Kidneys, Tumor, and Composite Dice, as outlined by the organizers. Kidneys Dice denotes the segmentation performance when considering both kidneys and tumors as the foreground, whereas Tumor Dice considers everything except the tumor as background. Composite Dice is simply the average of Kidneys Dice and Tumor Dice.

\section{Results and Discussion}

\newcommand\VRule[1][\arrayrulewidth]{\vrule width #1}

We evaluate the EG-CNN module when it is used to augment popular medical image segmentation models:  U-Net~\cite{Ronneberger15}, V-Net~\cite{Milletari16}, and Seg-Net~\cite{Myronenko18}.
We modified each architecture to adopt them to the given task and to be similar to the others for a more fair comparison. For both the U-net and V-net, we changed the normalization to Groupnorm, to better handle a small batch size, and adjusted the number of layers to a roughly equivalent number between the networks. For each dataset we trained the main CNN segmentation network with and without the EG-CNN in order to validate the contribution of our proposed module. We estimated the accuracy of each model in terms of Dice score for each class and of the overall average.

\paragraph{BraTS 2019:}

\begin{figure}[]
\begin{center}
\includegraphics[width=\x\linewidth,height=\x\linewidth]{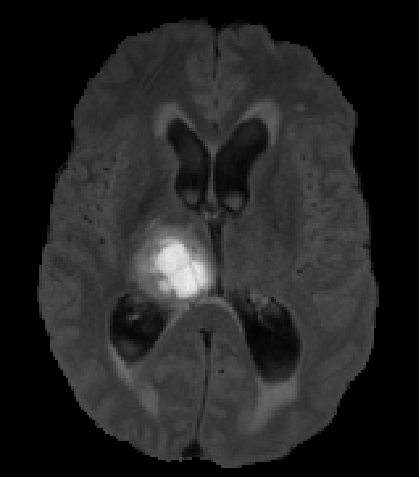}
\includegraphics[width=\x\linewidth,height=\x\linewidth]{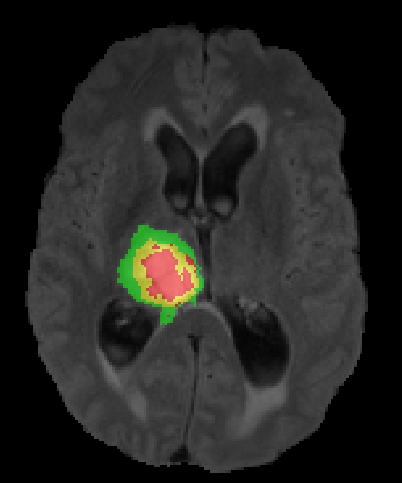}
\includegraphics[width=\x\linewidth,height=\x\linewidth]{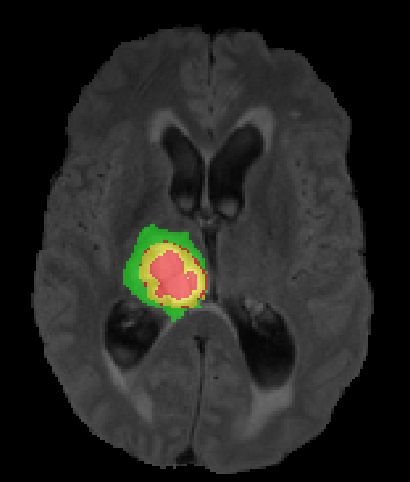}
\includegraphics[width=\x\linewidth,height=\x\linewidth]{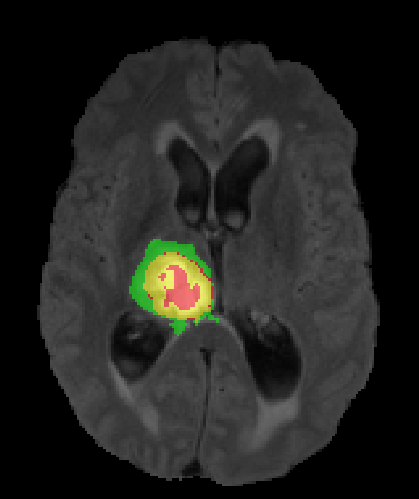}
\vspace{2pt}
\includegraphics[width=\x\linewidth,height=\x\linewidth]{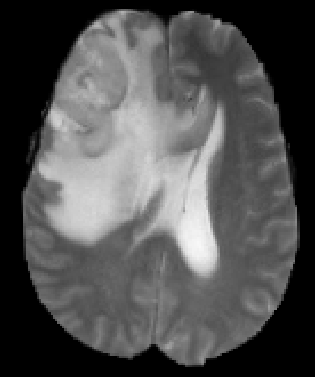}
\includegraphics[width=\x\linewidth,height=\x\linewidth]{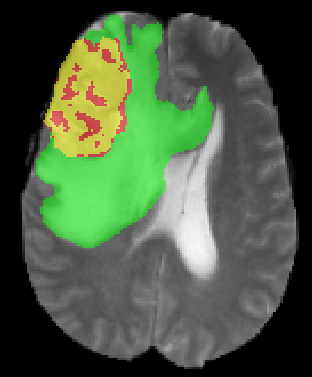}
\includegraphics[width=\x\linewidth,height=\x\linewidth]{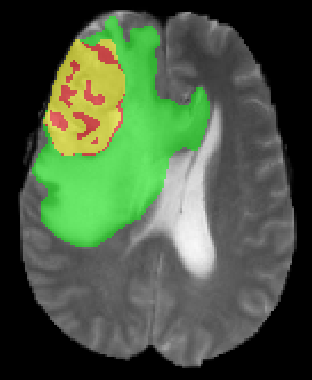}
\includegraphics[width=\x\linewidth,height=\x\linewidth]{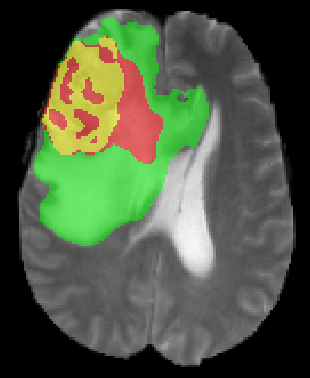}
\vspace{0.75pt}
\makebox[\x\linewidth]{(a) Input images} 
\makebox[\x\linewidth]{(b) Semantic Labels}
\makebox[\x\linewidth]{(c) Seg-Net+EG-CNN}
\makebox[\x\linewidth]{(d) Seg-Net}
\end{center}
\caption{In-plane visualization of the segmentation outputs for tumor sub-regions in BraTS 2019. Red, green and yellow labels denote TC, WT and ET sub-regions respectively.}
\label{fig:brats9}
\end{figure}

Table~\ref{table:result} reports the accuracy of the model on each of the classes: Whole Tumor (WT), Tumor Core (TC), and Enhancing Tumor (ET), as well as the overall overage accuracy. According to our benchmarks, including the EG-CNN consistently increases the overall and sub-region Dice scores in all cases. In the case of brain tumor segmentation, the EG-CNN has effectively learned highly complex and irregular boundaries of certain sub-regions. Therefore, it improves the segmentation quality around the edges, which leads to overall better segmentation performance. Figure~\ref{fig:brats9} illustrates how the addition of the EG-CNN to a standalone Seg-Net \cite{Myronenko18} improves the quality of segmentation. 

\begin{table*}[t]\setlength{\tabcolsep}{2pt}
\centering
\hspace{-3mm}\resizebox{1.0\linewidth}{!}{
\begin{tabular}{!{\VRule[2pt]}c!{\VRule[2pt]}c!{\VRule[2pt]}c!{\VRule[2pt]}c!{\VRule[2pt]}c!{\VRule[2pt]}c!{\VRule[2pt]}}

\specialrule{2pt}{0pt}{0pt}
\textbf{Architecture} & \textbf{Edge Stream}  &   \textbf{Average Dice} &  \textbf{ET  Dice}  &   \textbf{TC Dice}   &   \textbf{WT Dice}  \\\specialrule{2pt}{0pt}{0pt}
U-Net  & None   &  0.8305$\pm$0.0035&	0.7375$\pm$0.0021&	0.8480$\pm$0.0056& 0.9060$\pm$0.0021  \\
V-Net &None& 0.8281$\pm$0.0035&	0.7255$\pm$0.0049&	0.8570$\pm$0.0042& 0.9020$\pm$0.0014	 \\
Seg-Net &None&  0.8300$\pm$0.0033&0.7330$\pm$0.0042&0.8550$\pm$0.0049& 0.9015$\pm$0.0007\\
U-Net & EG-CNN  &  0.8406$\pm$0.0028&	0.7530$\pm$0.0113&	0.8630$\pm$0.0014& 0.9006$\pm$0.0042\\
V-Net &EG-CNN&  0.8386$\pm$0.0051&	0.7460$\pm$0.0056&	0.8605$\pm$0.0035& 0.9095$\pm$0.0063  \\
Seg-Net &EG-CNN&\textbf{0.8570$\pm$0.0007}&\textbf{0.7680$\pm$0.0113}&\textbf{0.8850$\pm$0.0070}&\textbf{0.9180$\pm$0.0028}
\\\specialrule{2pt}{0pt}{0pt}
\end{tabular}}
\bigskip
\caption{BraTS 2019 segmentation results in terms of overall and tumor sub-regions Dice scores. The Edge Stream column determines whether EG-CNN is utilized with the backbone architecture.}
\label{table:result}
\end{table*}

The quality of the predicted edges also validates the effectiveness of our proposed edge-aware loss function, since the boundaries are crisp and avoid the thickening effect around edges. Such a  phenomenon usually occurs when a naive loss function such as binary cross entropy is utilized for the task of edge prediction without taking precautions. Moreover, our model results in more fine-grained boundaries and visually attractive edges because the learned predicted boundaries are eventually fused with the final prediction output of the main encoder-decoder architecture.

Since the addition of the EG-CNN module increases the number of free parameters of the overall model, we have also experimented with larger standalone models (by increasing their depth and/or width), but doing so did not result in the better validation accuracy. This indicates that our module improves the overall segmentation accuracy not due to the model capacity increase, but due to the extra emphasis of edge information. 

\paragraph{KiTS 2019:}

\begin{figure*}[t]
\begin{center}
\includegraphics[width=\x\linewidth,height=\x\linewidth]{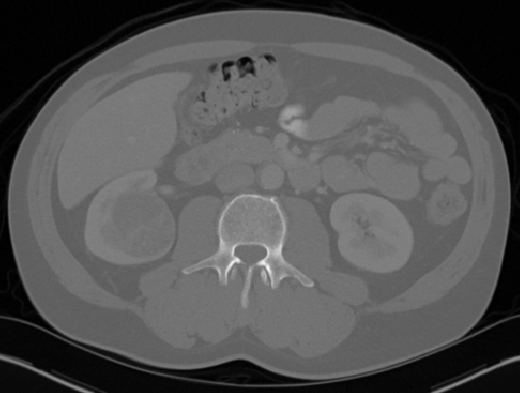}
\includegraphics[width=\x\linewidth,height=\x\linewidth]{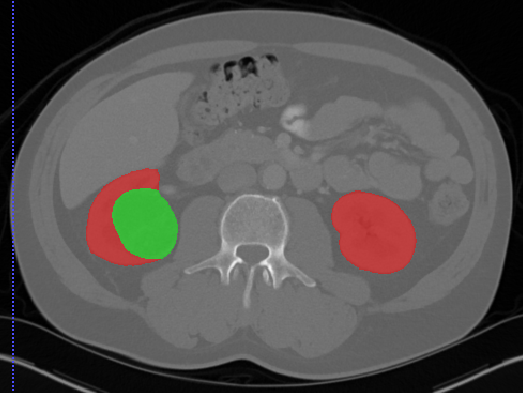}
\includegraphics[width=\x\linewidth,height=\x\linewidth]{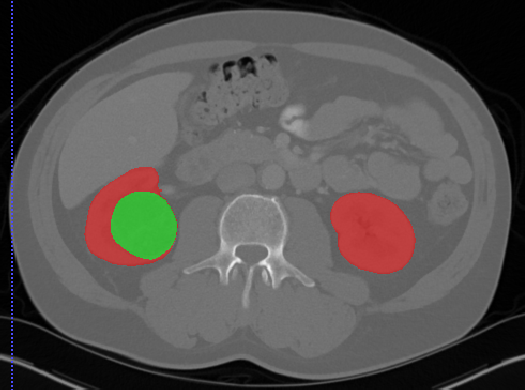}
\includegraphics[width=\x\linewidth,height=\x\linewidth]{{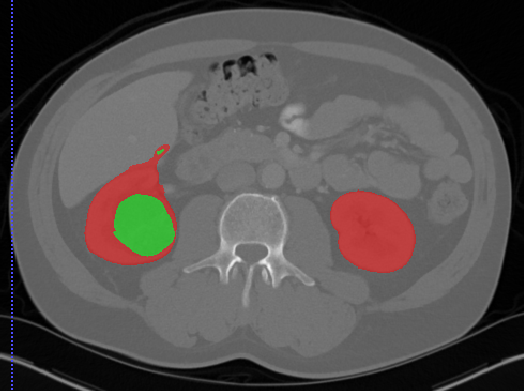}}

\vspace{2pt}

\includegraphics[width=\x\linewidth,height=\x\linewidth]{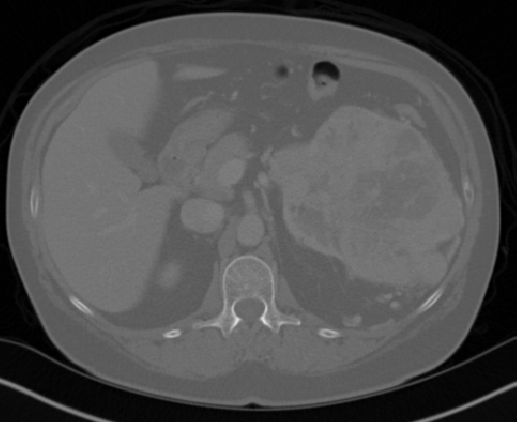}
\includegraphics[width=\x\linewidth,height=\x\linewidth]{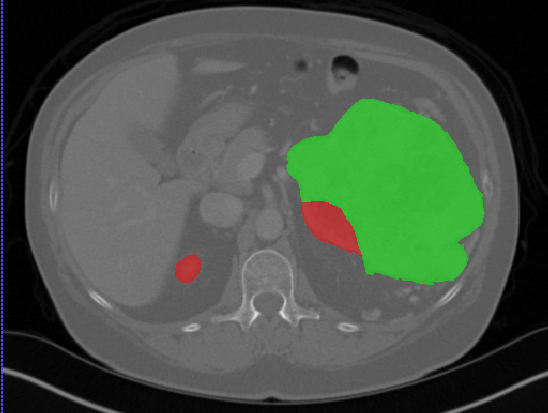}
\includegraphics[width=\x\linewidth,height=\x\linewidth]{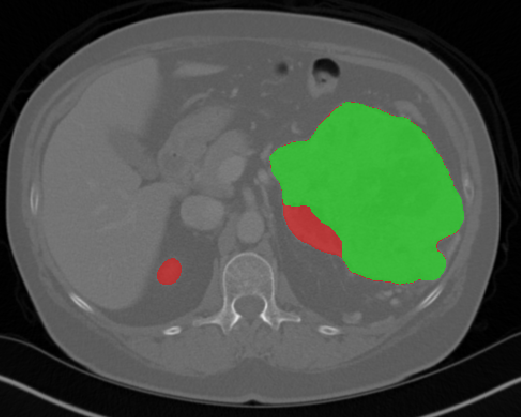}
\includegraphics[width=\x\linewidth,height=\x\linewidth]{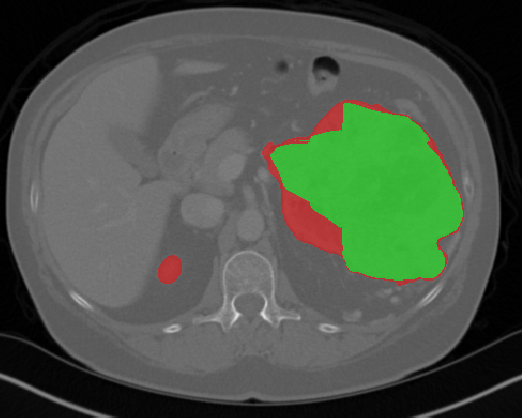}

\makebox[\x\linewidth]{(a) Input images} 
\makebox[\x\linewidth]{(b) Semantic Labels}
\makebox[\x\linewidth]{(c) Seg-Net+EG-CNN}
\makebox[\x\linewidth]{(d) Seg-Net}
\end{center}
\caption{Visualization of the segmentation performance of standalone Seg-Net and Seg-Net+EG-CNN on the KiTS 2019 challenge. The green and red masks denote tumor and kidneys, respectively.}
\label{fig:kits}
\end{figure*}

The achieved accuracy of the model for kidneys and kidney tumor classes, as well as the overall accuracy are presented in Table~\ref{table:kits}. Similar to the results achieved on BraTS 2019 dataset, the addition of EG-CNN has consistently improved the segmentation performance. Visual comparisons of the output segmentation and boundary predictions are presented in Figure~\ref{fig:kits}. As such, the predicted edges visually conform to the region outlines, demonstrating that the EG-CNN module and our proposed loss functions helped to captured the details of the edges. This has also been reflected in the final predictions of semantic masks.

\begin{table*}[t]\setlength{\tabcolsep}{2pt}
\centering
\hspace{-3mm}\resizebox{1.0\linewidth}{!}{
\begin{tabular}{!{\VRule[2pt]}c!{\VRule[2pt]}c!{\VRule[2pt]}c!{\VRule[2pt]}c!{\VRule[2pt]}c!{\VRule[2pt]}}

\specialrule{2pt}{0pt}{0pt}
\textbf{Architecture} & \textbf{Edge Stream}  &   \textbf{Kidneys Dice} &  \textbf{Tumor Dice}  &   \textbf{Composite Dice}  \\\specialrule{2pt}{0pt}{0pt}
U-Net& None    &  0.9515$\pm$0.0049&	0.8245$\pm$0.0091&	0.8880$\pm$0.0070	  \\
V-Net&None     &  0.9370$\pm$0.0065&	0.8072$\pm$0.0072&	0.8720$\pm$0.0068	  \\
Seg-Net&None     &  0.9530$\pm$0.0028&	0.8235$\pm$0.0049&	0.8892$\pm$0.0038	  \\
U-Net&  EG-CNN   &  0.9620$\pm$0.0056&	0.8270$\pm$0.0084&	0.8945$\pm$0.0070	  \\
V-Net&EG-CNN     &  0.9483$\pm$0.0048&	0.8275$\pm$0.0087&	0.8879$\pm$0.0067	  \\
Seg-Net& EG-CNN     &  \textbf{0.9647$\pm$0.0051}&	\textbf{0.8353$\pm$0.0025}&	\textbf{0.9000$\pm$0.0038}
\\\specialrule{2pt}{0pt}{0pt}
\end{tabular}}
\bigskip
\caption{KiTS 2019 segmentation results for kidneys, tumor, and composite Dice functions.}
\label{table:kits}
\end{table*}

\section{Conclusion}
\label{sec:conclusion}

We proposed the EG-CNN, a plug-and-play module for boundary-aware CNN segmentation, which can be paired with an existing encoder-decoder architecture to improve the segmentation accuracy. Our EG-CNN does not require any additional annotation effort since edge information can be extracted from the ground truth segmentation masks. Supervised by edge-aware and consistency loss functions, the EG-CNN learns to emphasize the edge representations by leveraging the feature maps of intermediate resolutions in the encoder of the main stream and feeding them into a series of edge-gated layers. We evaluated the EG-CNN by utilizing three popular 3D segmentation architectures, U-Net, V-Net, and Seg-Net, in the tasks of brain and kidney tumor segmentation on the BraTS19 and KiTS19 datasets. Our results indicate that the addition of the proposed EG-CNN consistently improves the segmentation accuracy in all the benchmarks.

\bibliography{midl-samplebibliography}






\end{document}